\documentclass[aps,prd,groupedaddress,showpacs,nofootinbib]{revtex4}

\usepackage{graphicx,dcolumn,bm,amssymb,amsmath,latexsym,amsfonts,footnote}

\begin{document}

\newcommand{\nonu}{\nonumber}
\newcommand{\sm}{\small}
\newcommand{\noi}{\noindent}
\newcommand{\npg}{\newpage}
\newcommand{\nl}{\newline}
\newcommand{\bp}{\begin{picture}}
\newcommand{\ep}{\end{picture}}
\newcommand{\bc}{\begin{center}}
\newcommand{\ec}{\end{center}}
\newcommand{\be}{\begin{equation}}
\newcommand{\ee}{\end{equation}}
\newcommand{\beal}{\begin{align}}
\newcommand{\eeal}{\end{align}}
\newcommand{\bea}{\begin{eqnarray}}
\newcommand{\eea}{\end{eqnarray}}
\newcommand{\bnabla}{\mbox{\boldmath $\nabla$}}
\newcommand{\univec}{\textbf{a}}
\newcommand{\VectorA}{\textbf{A}}
\newcommand{\Pint}

\title{Corotating dyonic binary black holes}

\author{I. Cabrera-Munguia\footnote{icabreramunguia@gmail.com}}
\affiliation{Departamento de F\'isica y Matem\'aticas, Universidad Aut\'onoma de Ciudad Ju\'arez, 32310 Ciudad Ju\'arez, Chihuahua, M\'exico}


\begin{abstract}
This paper is dedicated to derive and study binary systems of identical corotating dyonic black holes separated by a massless strut --two 5-parametric corotating binary black hole models endowed with both electric and magnetic charges-- where the dyonic black holes carrying equal/opposite electromagnetic charges in the first/second model satisfy the extended Smarr formula for the mass including the magnetic charge as a fourth conserved parameter.
\end{abstract}

\maketitle

\vspace{-0.4cm}
\section{Introduction}
\vspace{-0.3cm}
It is well-known that any black hole (BH) solution fulfilling the Einstein-Maxwell equations in stationary spacetimes can be described by only three conserved parameters: the mass, electric charge, and angular momentum \cite{MTW}. This statement is called the no-hair conjecture\cite{Israel0, CarterRobinson, Mazur} and a Kerr-Newman BH solution \cite{ENewman} is the one depicted by these physical parameters. Obviously, one may bear in mind the addition of the magnetic charge as a fourth conserved parameter since it is also conserved in Einstein-Maxwell theory \cite{Bekenstein}. In this context, a duality rotation (DR) studied long time ago by Carter \cite{Carter} seems to be the easiest path to add the magnetic charge into the Kerr-Newman solution in order to describe a rotating dyon --a particle containing both electric and magnetic charges \cite{Schwinger}--, which will satisfy an extended Smarr formula for the mass \cite{Smarr}. It should be mentioned that in this approach there exists no Dirac string (DS) (or monopole hair) joined to the BH.

On the other hand, Tomimatsu \cite{Tomi} via Komar integrals \cite{Komar} provided simple formulas that permits the derivation of the mass formula for multi-connected horizons. In particular, in a binary BH system he found that in the absence of the global net magnetic charge there appears a DS linking the BHs, in the way of magnetic flux. In this physical scenario the global magnetic charge is eliminated once each BH is equipped with an individual magnetic charge opposite in sign but carrying the same magnitude. Naturally that, the DS vanishes if there are no magnetic charges in the solution \cite{Tomi} recovering the Kerr-Newman BH description. It is worthwhile to stress the fact that there exist two approaches that allow us to add the magnetic charge into each BH; the DR \cite{Carter} and the inclusion of the DS into the mass \cite{Tomi,Galtsov}.

Following Tomimatsu's approach, in \cite{CLLM1,CLLM2,CLM} has been studied the contribution of the DS into the mass formula for two counterrotating dyonic BHs held apart by a massless strut \cite{BachW,Israel}, founding that each individual angular momentum suffers additional rotation provided by the presence of the DS, in the form $J-Q_{H}Q_{B}$. However, in a later paper published by Cl\'ement and Gal'tsov \cite{Galtsov} it was shown that Tomimatsu's formulas \cite{Tomi} are incorrect in the presence of both electric and magnetic charges, owing to the fact that the DS affects only the horizon mass and not the horizon angular momentum as was reported earlier in \cite{CLLM1,CLLM2,CLM}, where it is necessary to adopt a constant (or gauge) in the magnetic potential in order to give equal weights to balance the horizon mass and angular momentum.

The present paper has the main goal to use Carter's proposal \cite{Carter} on the DR to derive the dyonic extensions of two corotating Kerr-Newman binary BH models previously studied in \cite{CCHV}. These extended models are well represented by five physical arbitrary parameters: mass $M_{H}$, angular momentum $J_{H}$, electric charge $Q_{H}$, magnetic charge $B_{H}$, and a relative distance $R$, where all the thermodynamical properties contained inside the extended Smarr formula have been derived in a concise form. In addition, it is demonstrated that after the DR is applied, the physical Komar parameters $\{M_{H},Q_{H},B_{H},J_{H}\}$ are conserved without the need of adding any specific constant in the magnetic potential, contrary to the claim made in \cite{MankoCompean}. In this regard, is also added a short description on the correct use of this gauge with the aim to include the contribution of the DS in the horizon mass, by using once again the results of \cite{CLLM2} and the ones derived within this work for corotating systems.

\vspace{-0.4cm}
\section{Corotating dyonic binary black holes}
\vspace{-0.3cm}
Let us begin this section by introducing the following Ernst equations \cite{Ernst}:
\vspace{-0.1cm}
\bea \begin{split}  \left({\rm{Re}} {\cal{E}}+|\Phi|^{2}\right)\Delta{\cal{E}}&=(\bnabla{\cal{E}}+
2\bar{\Phi}\bnabla \Phi)\cdot \bnabla {\cal{E}}, \\
 \left({\rm{Re}}{\cal{E}}+|\Phi|^{2}\right)\Delta \Phi&=(\bnabla{\cal{E}}+
2\bar{\Phi}\bnabla\Phi)\cdot\bnabla\Phi, \label{Ernst} \end{split} \eea

\vspace{-0.1cm}
\noi that are equivalent to Einstein-Maxwell field equations in stationary axisymmetric spacetimes, where $({\cal{E}},\Phi)$ are complex potentials given by ${\cal{E}}=f-|\Phi|^{2}+i\Psi$ and $\Phi=-A_{4}+i A'_{3}$. We notice that Eq.\ (\ref{Ernst}) remain invariant under DR, in which $\Phi$ might be replaced by $\Phi e^{i \alpha}$, where $\alpha$ is a constant duality angle closely linked to the magnetic charge. On the other hand, the line element defining stationary axisymmetric spacetimes is given by \cite{Papapetrou}
\vspace{-0.1cm}
\be ds^{2}=f^{-1}\left[e^{2\gamma}(d\rho^{2}+dz^{2})+\rho^{2}d\varphi^{2}\right]- f(dt-\omega d\varphi)^{2},
\label{Papapetrou}\ee

\vspace{-0.1cm}
\noi where $f(\rho,z)$,$\omega(\rho,z)$, and $\gamma(\rho,z)$ are metric functions that can be obtained once one is able to solve following differential equations
\vspace{-0.2cm}
\begin{align}
4\gamma_{,\rho}&=\rho f^{-2} \left[|{\cal{E}}_{,\rho}+
2\bar{\Phi}\Phi_{,\rho}|^{2} -|{\cal{E}}_{,z}+ 2\bar{\Phi}\Phi_{,z}|^{2}\right] - 4\rho f^{-1}(|\Phi_{,\rho}|^{2}- |\Phi_{,z}|^{2}),\nonu\\
2\gamma_{,z}&=\rho f^{-2}{\rm{Re}} \left[({\cal{E}}_{,\rho}+
2\bar{\Phi}\Phi_{,\rho})(\bar{{\cal{E}}}_{,z}+ 2\bar{\Phi}\Phi_{,z})\right]-4\rho f^{-1} {\rm{Re}(\bar{\Phi}_{,\rho}\Phi_{,z})},\nonu\\
\omega_{,\rho}&=-\rho f^{-2}{\rm{Im}}( {\cal{E}}_{,z}+ 2 \Phi\bar{\Phi}_{,z}),\qquad
\omega_{,z}=\rho f^{-2}{\rm{Im}}( {\cal{E}}_{,\rho}+ 2 \Phi\bar{\Phi}_{,\rho}).
\label{metricfunctions}\end{align}

\vspace{-0.0cm}
Then we have that the explicit knowledge of the Ernst potentials $({\cal{E}},\Phi)$ fulfilling Eq.\ (\ref{Ernst}), will be helpful to solve Eq.\ (\ref{metricfunctions}). In order to derive the Ernst potentials as well as the metric functions in the entire space $(\rho,z)$ one can make use of Sibgatullin's method \cite{Sibgatullin}, where it is necessary to adopt first a specific form of the Ernst potentials on the symmetry axis (the axis data). In \cite{CCHV} we have used the following axis data for asymptotically flat spacetimes
\vspace{-0.1cm}
\begin{align}
{\cal E}(0,z)&=\frac{z^{2}-2(M + i\mathfrak{q})z +2\Delta-R^{2}/4-\sigma^{2}-2q_{o}(Q/M)+i\delta}{z^{2}+ 2(M -i\mathfrak{q})z + 2\Delta-R^{2}/4-\sigma^{2}+2q_{o}(Q/M)-i\delta}, \nonu\\
\Phi(0,z)&=\frac{2(Qz+\mathfrak{q}_{o})}{z^{2}+ 2(M -i\mathfrak{q})z + 2\Delta-R^{2}/4-\sigma^{2}+2q_{o}(Q/M)-i\delta},\nonu\\
\sigma&=\sqrt{\Delta- \frac{4\big[|\mathfrak{q}_{o}|^{2}-(Q/M)^{2}q_{o}^{2}\big]-\delta^{2}}{R^{2}-4\Delta}}, \qquad
\Delta= M^{2}-Q^{2}-\mathfrak{q}^{2},  \qquad \mathfrak{q}_{o}=q_{o}+ib_{o},
\label{ernstaxiselectroI}\end{align}

\begin{figure}[ht]
\centering
\includegraphics[width=6.0cm,height=5.0cm]{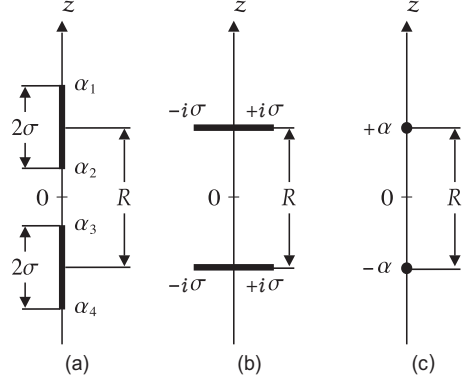}
\vspace{-0.1cm}
\caption{Identical Kerr-Newman-type sources on the symmetry axis: (a) BHs for $\sigma^{2}>0$; (b) hyperextreme sources if $\sigma \rightarrow i \sigma$ (or $\sigma^{2}<0$ ); (c) extreme BHs where $\sigma=0$.}
\label{DK}\end{figure}

\vspace{-0.1cm}
\noi with the main objective to describe corotating binary systems of identical Kerr-Newman BHs \cite{ENewman}. From the aforementioned Eq.\ (\ref{ernstaxiselectroI}) after applying the Hoenselaers-Perj\'es procedure \cite{HP,Sotiriou}, it is possible to obtain the first Simon's multipoles \cite{Simon}, where the total mass, total electric charge, and total angular momentum of the binary system are represented by $2M$, $2Q$, and $4M\mathfrak{q}-\delta$, respectively, while the total electromagnetic dipole moment is given as $2\big[q_{o}+i(b_{o}+2\mathfrak{q}Q)\big]$. Moreover, $R$ defines a separation distance between the sources (see Fig.\ \ref{DK}), which may be BHs if $\sigma^{2}\geq 0$ or hyperextreme sources when $\sigma^{2}<0$. In this framework, the Ernst potentials, Kinnersley potential $\Phi_{2}$ \cite{Kinnersley} as well as the metric functions in the whole spacetime were obtained in Ref.\ \cite{CCHV}; these explicitly are
\vspace{-0.0cm}
\begin{align}
{\cal{E}}&=\frac{\Lambda+\Gamma}{\Lambda-\Gamma},\quad \Phi=\frac{\chi}{\Lambda-\Gamma},\qquad \Phi_{2}=\frac{F}{\Lambda-\Gamma},\qquad
f=\frac{|\Lambda|^{2}-|\Gamma|^{2}+ |\chi|^{2}}{|\Lambda-\Gamma|^{2}},\qquad \omega=4\mathfrak{q}+\frac{{\rm{Im}}\left[(\Lambda-\Gamma)\overline{\mathcal{G}}-\chi \overline{\mathcal{I}} \right]}{|\Lambda|^{2}-|\Gamma|^{2}+ |\chi|^{2}},\nonu\\
e^{2\gamma}&=\frac{|\Lambda|^{2}-|\Gamma|^{2}+ |\chi|^{2}}{64\sigma^{4}R^{4}\kappa_{o}^{2} r_{1}r_{2}r_{3}r_{4}}, \qquad
\Lambda=2\sigma^{2} \left[R^{2}\kappa_{o}(r_{1}+r_{2})(r_{3}+r_{4})+4a(r_{1}-r_{3})(r_{2}-r_{4})\right]\nonu\\ &+2R^{2}\left[\kappa_{o}(2\Delta-\sigma^{2})-a\right](r_{1}-r_{2})(r_{3}-r_{4})
+2iR\bigg\{\Big(2\mathfrak{q}{\rm Re}(s_{+})+{\rm Im}(p_{+})\Big)\Big[R(\mathfrak{r}_{1}-\mathfrak{r}_{2})(r_{3}-r_{4})\nonu\\
&-2\sigma\big(\mathfrak{r}_{1}r_{4}-\mathfrak{r}_{2}r_{3}+4\sigma r_{3}r_{4}\big)\Big] +\mathfrak{q}\kappa_{o}\Big[ r_{1}\big( R^{2}r_{3}-\kappa_{o}r_{4}\big)-r_{2}\big(\kappa_{o} r_{3}-R^{2}r_{4}\big)-8\sigma^{2}r_{3}r_{4} \Big]\bigg\},\nonu\\
\Gamma&=4\sigma R\left(M\Gamma_{o}- b\chi_{+}\right),\quad F=(4\mathfrak{q}+iz)\chi-i\mathcal{I},\quad \chi=-4\sigma R\left(Q\Gamma_{o}+2\mathbb{Q}\chi_{+}\right),\quad \Gamma_{o}=R \chi_{-}-2\sigma \chi_{s}+2\chi_{1+}, \nonu\\
\mathcal{G}&= 2z\Gamma+  8\sigma^{2}\bigg\{ R\Big[2\big({\rm Re}(a)-2|\mathfrak{q}_{o}|^{2}\big)
+Q^{2}\kappa_{o}\Big](r_{1}r_{2}-r_{3}r_{4})+2i\mathfrak{q}R^{2}\kappa_{o}(r_{2}r_{3}+r_{1}r_{4})+2i\Big[R{\rm Im}(a)+Q\xi_{0}-4\mathfrak{q}|\mathfrak{q}_{o}|^{2}\Big] \nonu\\
&\times (r_{1}-r_{3})(r_{2}-r_{4})\bigg\}-4R^{2}\bigg\{\sigma\left[2a-(R-2\sigma)
\Big(2(R+2i\mathfrak{q})s_{+}+p_{+} \Big)\right]+i\left(Q\xi_{o}+2Qb_{o}\kappa_{o}-4\mathfrak{q}|\mathfrak{q}_{o}|^{2}\right)\bigg\}
(r_{1}-r_{2})(r_{3}-r_{4})\nonu\\
&+2\sigma R\bigg\{4R\Big(2\kappa_{o} \Delta -{\rm Re}(a)\Big)r_{4}+\Big[Q(4q_{o}+QR)\kappa_{o}+4R|\mathfrak{q}_{o}|^{2}\Big](r_{3}+r_{4})\bigg\}(r_{1}-r_{2})\nonu\\
&+2\sigma R \bigg\{4R\Big(2\kappa_{o} \Delta -{\rm Re}(a)\Big)r_{2}-\Big[Q(4q_{o}-QR)\kappa_{o}-4R|\mathfrak{q}_{o}|^{2}\Big](r_{1}+r_{2})\bigg\}(r_{3}-r_{4})
+4M\sigma R\big(\kappa_{o}\chi_{+} +2R\chi_{1-}+4\sigma\chi_{p}\big)\nonu\\
&-4b\sigma R(R\chi_{-}+2\sigma\chi_{s})-8\sigma R(Qb+2M\mathbb{Q})\Big[ 2\bar{\mathfrak{q}}_{o}\big(\mathfrak{r}_{1}-\mathfrak{r}_{2}+\mathfrak{r}_{3}-\mathfrak{r}_{4}\big)
+Q\kappa_{o}(r_{1}-r_{2}-r_{3}+r_{4})\Big],\nonu
\end{align}

\begin{align}
\mathcal{I}&=A\Big[4\sigma^{2}(r_{1}-r_{3})(r_{2}-r_{4})-R^{2}(r_{1}-r_{2})(r_{3}-r_{4})\Big]
+R\kappa_{-}\Big[B_{+}\kappa_{o}r_{1}-B_{-}R\mathfrak{r}_{2} \Big]r_{4}+R\kappa_{+}\Big[B_{-}\kappa_{o}r_{2}-B_{+}R\mathfrak{r}_{1} \Big]r_{3}\nonu\\
&-16\sigma^{2}R \Big\{\Big[M(R+2\sigma)(\kappa_{+}+2QR)-B_{+}\mathfrak{q}_{o}\Big]r_{3}r_{4}-R\kappa_{o}(2M\mathbb{Q}+Qb)\Big\}+
8\mathbb{Q}\sigma R(\chi_{1+}+\sigma\chi_{s})\nonu\\
&+ 2\sigma R\Big[Q\big(2R^{2}-8\Delta+\kappa_{o}\big)+8i\mathfrak{q} \mathbb{Q}\Big]\chi_{+}+ 12\sigma R^{2}\mathbb{Q}\chi_{-} + 8Q\sigma R(R\chi_{1-}+2\sigma\chi_{p}), \quad \kappa_{o}=R^{2}-4\sigma^{2}, \nonu\\
\chi_{\pm}&=s_{+}\mathfrak{r}_{1}-s_{-}\mathfrak{r}_{2} \pm
(\bar{s}_{-}\mathfrak{r}_{3}-\bar{s}_{+}\mathfrak{r}_{4}) , \quad \chi_{1\pm}=p_{+}\mathfrak{r}_{1}+p_{-}\mathfrak{r}_{2}\pm
(\bar{p}_{-}\mathfrak{r}_{3}+\bar{p}_{+}\mathfrak{r}_{4}),\quad
\chi_{s}=s_{+}\mathfrak{r}_{1}+s_{-}\mathfrak{r}_{2}+\bar{s}_{-}\mathfrak{r}_{3}+\bar{s}_{+}\mathfrak{r}_{4}, \nonu\\
\chi_{p}&=p_{+}\mathfrak{r}_{1}-p_{-}\mathfrak{r}_{2} + \bar{p}_{-}\mathfrak{r}_{3}-\bar{p}_{+}\mathfrak{r}_{4},\quad
a=(R+2i\mathfrak{q})p_{+}-s_{+}\big[s_{+}-(R+2i\mathfrak{q})^{2}\big], \quad
b=-2q_{o}(Q/M)+i(\delta-4M\mathfrak{q}),\nonu\\
A&=4M\Big[ \Big( 2 \mathbb{Q}+ Q(R-2\sigma)\Big)s_{+}+2Qp_{+}\Big]
+B_{+}\Big[Q\big(R^{2}-4\Delta\big)-2(R+2i\mathfrak{q})\mathfrak{q}_{o} \Big], \quad \kappa_{\pm}=2\mathfrak{q}_{o}-Q(R\pm 2\sigma), \nonu\\
B_{\pm}&=\Big[ R s_{\pm} \pm p_{\pm} +2Q\big(2\bar{\mathfrak{q}}_{o}+Q(R\pm 2\sigma)\big)\Big]/M, \quad
p_{\pm}=-\sigma(R^{2}-4\Delta)\pm i\big[2M\delta+4b_{o}Q-(R+2i\mathfrak{q}){\rm Im}(s_{\pm})\big], \nonu\\ s_{\pm}&=2\Delta \pm \sigma R+ i \mathfrak{q}(R\pm 2\sigma),\quad
\xi_{o}=4Q\Big[M \delta+2b_{o} Q+\mathfrak{q}(\Delta-\sigma^{2})\Big]-(2b_{o}+\mathfrak{q} Q)(R^{2}-4\Delta), \quad \mathbb{Q}=\mathfrak{q}_{o}+2i \mathfrak{q} Q,\nonu\\
\mathfrak{r}_{1,4}&=(R-2\sigma)r_{1,4}, \quad
\mathfrak{r}_{2,3}=(R+2\sigma)r_{2,3}, \quad r_{1,2}=\sqrt{\rho^{2}+\left(z-R/2 \mp \sigma\right)^{2}}, \quad
r_{3,4}=\sqrt{\rho^{2}+\left(z+R/2 \mp \sigma\right)^{2}},
\label{sevenparametersI}\end{align}

\vspace{-0.1cm}
\noi where $r_{n}=\sqrt{\rho^{2}+(z-\alpha_{n})^{2}}$ are the distances from the value $\alpha_{n}$ defining the location of the source to any point $(\rho,z)$ outside the symmetry axis as shown in Fig.\ \ref{DK}. On the other hand, it is necessary to impose the next axis condition
\vspace{-0.1cm}
\be \omega\Big(\rho=0, |z|< {\rm{Re}}(\alpha_{2})\Big)=0,\label{omegamiddle}\ee

\vspace{-0.1cm}
\noi with the aim to disconnect the region in between sources. Once is used the metric function $\omega$ from Eq.\ (\ref{sevenparametersI}) such a condition can be represented by a very simple quadratic equation for solving
\vspace{-0.1cm}
\begin{align} &8 \mathfrak{q} P_{0}b_{o}^{2}+2P_{0}(2Q b_{o}+M \delta)(R^{2}-4\Delta)-\big[2\mathfrak{q}s_{o}-(R+2M)\delta\big] (R^{2}-4\Delta)^{2}+4\mathfrak{q}\Big\{(P_{0}-2s_{o})\Big[2q_{o}^{2}\big(1-2(Q/M)^{2}\big)-\delta^{2}\Big]\nonu\\
&+4s_{o}q_{o}^{2}\Big\}=0, \nonu\\
P_{0}&=(R+2M)^{2}+4\mathfrak{q}^{2}, \quad s_{o}=M(R+2M)-Q^{2}.
\label{middle1}\end{align}

\vspace{-0.1cm}
On the other hand, Eq.\ (\ref{sevenparametersI}) permits us to calculate the physical Komar parameters \cite{Komar} for BHs via the amended Tomimatsu formulae \cite{Tomi,Galtsov}
\vspace{-0.1cm}
\begin{align} M_{H}&= -\frac{1}{8\pi}\int_{H} \omega \Psi_{,z}\, d\varphi dz-M_{A}^{S}, \qquad
Q_{H}=\frac{1}{4\pi}\int_{H}\omega A_{3,z}^{'}\, d\varphi dz, \qquad B_{H}=\frac{1}{4\pi }\int_{H}\omega A_{4,z}\, d\varphi dz,\nonu\\
J_{H}&=-\frac{1}{8\pi}\int_{H}\omega\left[1+ \frac{\omega \Psi_{,z}}{2}
-\tilde{A}_{3}A_{3,z}^{'}\right]d\varphi dz -\frac{\omega^{H}M_{A}^{S}}{2},& \label{TomiGaltsov} \end{align}

\vspace{-0.1cm}
\noi where $\omega^{H}$ is the constant value for $\omega$ at the horizon while $\tilde{A}_{3}=A_{3}+\omega A_{4}$, being $A_{3}$ the magnetic potential obtainable from the real part of $\Phi_{2}$ in the following way:
\vspace{-0.1cm}
\be A_{3}={\rm Re}(\Phi_{2})=-4\mathfrak{q} A_{4}-zA'_{3}+{\rm Im} \bigg(\frac{\mathcal{I}}{\Lambda-\Gamma}\bigg). \ee

\vspace{-0.1cm}
Furthermore $A_{4}=-{\rm {Re} (\Phi)}$, and $M_{A}^{S}$ is a boundary term related to the presence of the DS connecting the BHs that is computed by virtue of
\vspace{-0.1cm}
\be M_{A}^{S}=-\frac{1}{4\pi}\int_{H}(A_{3}^{'}A_{3})_{,z}d\varphi dz.\ee

\vspace{-0.1cm}
The expressions contained in Eq.\ (\ref{TomiGaltsov}) can be rearranged to derive the Smarr formula \cite{Smarr} for the horizon mass of each BH \cite{Galtsov},
\vspace{-0.1cm}
\be M_{H}=\frac{\kappa S}{4\pi} +2 \Omega J_{H} +\Phi^{H}_{E} Q_{H}=\sigma +2 \Omega J_{H} +\Phi^{H}_{E} Q_{H}, \label{Smarr}\ee

\vspace{-0.1cm}
\noi where $\Omega=1/\omega^{H}$ is the angular velocity and $\Phi^{H}_{E}=-A_{4}^{H}-\Omega A_{3}^{H}$ is the electric potential evaluated on the horizon. It is worth noting that we have written down the formula for the mass Eq.\ (\ref{Smarr}) with two different aspects since the area of the horizon $S$ and surface gravity $\kappa$ are related to the half-length horizon $\sigma$ by means of \cite{Tomi,Carter2}
\vspace{-0.1cm}
\be S= \frac{4\pi \sigma}{\kappa}, \qquad \kappa=\sqrt{-\Omega^{2}e^{-2\gamma^{H}}},\ee

\vspace{-0.3cm}

\vspace{-0.1cm}
\noi where $\gamma^{H}$ is the value acquired by the metric function $\gamma$ over the horizon. Due to the fact that each thin rod representing the BH horizon in Fig.\ \ref{DK}(a) contains the same length, without loss of generality, it is possible to calculate the Komar parameters by means of the values $H=\{-\sigma\leq z-R/2 \leq\sigma, 0\leq \varphi \leq 2\pi, \rho \rightarrow 0\}$, that define the upper BH. After using the Ernst potentials and metric functions derived in Ref.\ \cite{CCHV}, the horizon mass $M_{H}$ and electromagnetic charge $Q_{H}+iB_{H}$ assume the form
\vspace{-0.1cm}
\begin{align} & M_{H}=M+ \frac{2q_{o}(Q/M)P_{0} R(R^{2}-4\Delta)}{\Big[(R+2M)(R^{2}-4\Delta)-4\mathfrak{q}\delta\Big]^{2}+64\mathfrak{q}^{2}q_{o}^{2}(Q/M)^{2}}
-M_{A}^{S},\nonu\\
&Q_{H}+i B_{H}= Q  +2\frac{P_{0}(q_{o}+ib_{o})+iQ\Big(\mathfrak{q}(R^{2}-4\Delta)+(R+2M)\big[\delta+2iq_{o}(Q/M)\big]\Big)}
{(R+2M)(R^{2}-4\Delta)-4\mathfrak{q}\big[\delta+2iq_{o}(Q/M)\big]},
\label{totalcharge}\end{align}

\vspace{-0.1cm}
\noi whereas the lower BH defines its corresponding horizon mass once is performed the change $q_{o}\rightarrow -q_{o}$ in Eq.\ (\ref{totalcharge}), while its electromagnetic charge is $2Q-Q_{H}-iB_{H}$. In Ref.\ \cite{CCHV}, the axis condition Eq.\ (\ref{middle1}) has been combined together with Eq.\ (\ref{totalcharge}) in order to derive two corotating electrically charged models; i.e., two binary systems of equal corotating Kerr-Newman BHs endowed with identical or opposite electric charges and separated by a massless strut, in which a DS binding the BHs \emph{does not exist} since the absence of individual magnetic charges was also established. However, the contribution of the DS into the horizon mass might be considered if $B_{H} \neq 0 $, where after setting $q_{o}=0$ in Eqs.\ (\ref{middle1}) and (\ref{totalcharge}) eventually one gets the result
\vspace{-0.1cm}
\begin{align}  \delta&=\frac{2(R^{2}-4\Delta)\big[M \mathfrak{q} P_{0}+(R+2M) \tau \big]}{(R^{2}+2MR+4\mathfrak{q}^{2})P_{0}
+8\mathfrak{q} \tau},\qquad
b_{o}=-\frac{(R^{2}-4\Delta)\big((2\mathfrak{q}Q_{H}-B_{H}R)P_{0}+4Q_{H}\tau \big)}{2\big[(R^{2}+2MR+4\mathfrak{q}^{2})P_{0}
+8\mathfrak{q} \tau\big]}, \nonu\\
\tau&=\mathfrak{q}(Q_{H}^{2}-B_{H}^{2})-B_{H}Q_{H}(R+2M), \qquad Q=Q_{H},
\label{Sol1}\end{align}

\vspace{-0.1cm}
\noi that permits the description of a two-body system of corotating dyonic BHs connected by a DS, where the upper and lower constituents have electromagnetic charges equal to $Q_{H}+iB_{H}$ and $Q_{H}-iB_{H}$ respectively. Lengthy calculations eventually give us simple expressions for $M_{A}^{S}$, $\Phi_{E}^{H}$, and $\Omega$,  namely
\vspace{-0.1cm}
\begin{align}
M_{A}^{S}&= B_{H} \epsilon \Bigg(\frac{B_{H}\mathcal{L}-Q_{H}\mathcal{M}}{\mathcal{L}^{2}+\mathcal{M}^{2}}\Bigg), \qquad
\Phi_{E}^{H}= \frac{Q_{H}(R+2\sigma)\mathcal{L}-2b_{0}\mathcal{M}}{\mathcal{L}^{2}+\mathcal{M}^{2}},\qquad
\Omega=\frac{2\mathfrak{q}}{P_{0}}+ \frac{\epsilon\mathcal{M}}{\mathcal{L}^{2}+\mathcal{M}^{2}}, \nonu\\ \epsilon&=\frac{P_{0}R(R^{2}-4\Delta)}{(R^{2}+2MR+4\mathfrak{q}^{2})P_{0}
+8\mathfrak{q} \tau},\qquad
\mathcal{L}=MR+2\Delta+(R+2M)\sigma, \qquad \mathcal{M}= \delta+\mathfrak{q}(R+2\sigma).
\label{ThermoDS1}\end{align}

\vspace{-0.1cm}
A second model emerges immediately from Eqs.\ (\ref{middle1}) and (\ref{totalcharge}) by doing first $Q=0$, to obtain
\vspace{-0.1cm}
\begin{align}
\delta&=\frac{2\mathfrak{q}(R^{2}-4\Delta_{1})\big[MP_{0}-|\mathcal{Q}|^{2}(R+2M)\big]}
{(R^{2}+2MR+4\mathfrak{q}^{2})P_{0}
-8\mathfrak{q}^{2}|\mathcal{Q}|^{2}},  \qquad
\mathfrak{q}_{o}=\frac{\mathcal{Q}R(R^{2}-4\Delta_{1})P_{0}}{2\Big[(R^{2}+2MR+4\mathfrak{q}^{2})P_{0}
-8\mathfrak{q}^{2}|\mathcal{Q}|^{2}\Big]}, \nonu\\
\Delta_{1}&=M^{2}-\mathfrak{q}^{2}, \qquad \mathcal{Q}=Q_{H}+iB_{H}, \qquad |\mathcal{Q}|^{2}=Q_{H}^{2}+B_{H}^{2}, \label{Sol2}\end{align}

\vspace{-0.1cm}
\noi thus having a binary system that contains opposite electromagnetic charges, where now it is possible to get
\vspace{-0.1cm}
\begin{align}
M_{A}^{S}&= 2B_{H}\Bigg(  \frac{b_{o}\mathcal{N}-q_{o}\mathcal{M}}{\mathcal{N}^{2}+\mathcal{M}^{2}}-\frac{2\mathfrak{q}Q_{H}}{P_{0}}\Bigg),
\qquad \Phi_{E}^{H}= 2\frac{q_{o}\mathcal{N}-b_{o}\mathcal{M}}{\mathcal{N}^{2}+\mathcal{M}^{2}}, \qquad
\Omega=\frac{2\mathfrak{q}}{P_{0}}+ \frac{2(q_{o}/Q_{H})\mathcal{M}}{\mathcal{N}^{2}+\mathcal{M}^{2}}, \nonu\\
\mathcal{N}&= MR+2\Delta_{1}+ (R+2M)\sigma.
\label{ThermoDS2}\end{align}

\vspace{-0.1cm}
In both models, it should be observed the contribution of the DS via the magnetic charge $B_{H}$, where regardless of which one may be considered, the upper and lower dyonic BHs are equipped with a magnetic charge $B_{H}$ and $-B_{H}$ respectively. That each source carries a magnetic charge of opposite sign is one of the main characteristics of dyonic binary BH models joined by a DS, when the global monopolar magnetic charge has been eliminated from the configuration \cite{CCHV}. It is important to comment that both dyonic BHs having a DS in between sources will satisfy the Smarr formula Eq.\ (\ref{Smarr}), in which their horizon mass is $M_{H}=M-M_{A}^{S}$.

\vspace{-0.4cm}
\subsection{Physical meaning of the DS into the horizon mass}
\vspace{-0.3cm}
At this point, we would like to discuss a little bit on some physical implications of $M_{A}^{S}$ within the Smarr formula. As has been recently proved by Cl\'ement and Gal'tsov \cite{Galtsov}, the correctness of Tomimatsu's formulas fails at the moment of including magnetic charges, due mainly to the fact that the horizon mass $M_{H}$ in Tomimatsu's approach does not contain the extra component $M_{A}^{S}$. For that reason Tomimatsu's expression for the horizon mass is determined by \cite{Tomi}
\vspace{-0.1cm}
\be M_{H}=\sigma +2 \Omega J_{H} +\Phi^{H}_{E} Q_{H}+M_{A}^{S}, \label{SmarrTomi}\ee

\vspace{-0.1cm}
\noi which looks different than the one displayed in Eq.\ (\ref{Smarr}). However, if we want to consider the contribution of the DS provided by the extra term $M_{A}^{S}$ into the horizon mass we need to include a constant\footnote{This constant gives a symmetric weight for each DS joined to the BH, providing a balance of mass and angular momentum. For a single dyonic Kerr-Newman BH the magnetic potential might be written as $A_{3new}=K_{0}+B_{H}(1-y)-a(1-y^{2})A_{4}$, where $a \equiv J_{H}/M$ and $(x,y)$ are prolate coordinates. The choice $K_{0}=-B_{H}$ is equivalent to carry out $C=0$ as was done in Ref.\ \cite{Galtsov}.} in the magnetic potential, namely
\vspace{-0.1cm}
\be A_{3 new}=K_{0}+A_{3},\ee

\vspace{-0.1cm}
\noi and it creates changes in the terms $M_{A}^{S}$ and $\Phi_{E}^{H}$ in the form
\vspace{-0.1cm}
\begin{align} M_{A new}^{S}&=-\frac{1}{4\pi}\int_{H}(A_{3}^{'}A_{3 new})_{,z}d\varphi dz=M_{A}^{S}-K_{0}\Omega Q_{H},\nonu\\
\Phi_{E new}^{H}&=-A_{4}^{H}-\Omega A_{3new}^{H}=\Phi_{E}^{H}-K_{0}\Omega.\label{thenewmass}
\end{align}

\vspace{-0.1cm}
In order to explain how this constant might be used, and with the purpose to amend the discrepancy committed in \cite{CLLM1,CLLM2,CLM}, we are going to appeal first to the results of the paper \cite{CLLM2} concerning to an oppositely electromagnetic charged two-body system of identical counterrotating BHs, where its thermodynamical properties were explicitly calculated; they read
\vspace{-0.1cm}
\begin{align} M_{A}^{S}&=B_{H}(B_{H}\phi^{H}-Q_{H}\Omega), \qquad \Phi_{E}^{H}=Q_{H}\phi^{H}-B_{H}\Omega, \nonu\\
\Omega&=\frac{\mu}{2}\frac{(R+2\sigma)\sqrt{X-1}}{M[R+2\sigma-(R-2M)X]-\mu |\mathcal{Q}|^{2}},\qquad
\phi^{H}=\frac{\mu}{2}\frac{R+2\sigma-(R-2M)X}{M[R+2\sigma-(R-2M)X]-\mu |\mathcal{Q}|^{2}},\nonu\\
\sigma&=\sqrt{X\big(M^{2}-|\mathcal{Q}|^{2}\mu\big)+\frac{R^{2}}{4}(1-X)}, \quad
\mu=\frac{R-2M}{R+2M},\label{counter}
\end{align}

\vspace{-0.1cm}
\noi where $X$ is an auxiliary variable defined as
\vspace{-0.1cm}
\be X=\frac{q_{o}^{2}+b_{o}^{2}}{(R/2-M)^{2}(Q_{H}^{2}+ B_{H}^{2})}, \label{theX}\ee

\vspace{-0.1cm}
\noi which in the lack of rotation turns out to be equal to the unit. Due to the fact that the mass formula Eq.\ (\ref{Smarr}) now takes the form
\vspace{-0.1cm}
\be M_{H}=M-M_{A new}^{S}=\sigma + 2\Omega J_{H} +\Phi^{H}_{E new} Q_{H}, \label{Smarr2}\ee

\vspace{-0.1cm}
\noi it is quite natural to deduce from Eqs.\ (\ref{thenewmass})-(\ref{counter}) that $K_{0}$ may be chosen as $K_{0}=-B_{H}$. Therefore, Eq.\ (\ref{Smarr2}) is further simplified as follows
\vspace{-0.1cm}
\be M_{H}\equiv M-B_{H}^{2}\phi^{H}=\sigma + 2\Omega J_{H} +Q_{H}^{2}\phi^{H}. \label{Smarr3}\ee

\vspace{-0.1cm}
The substitution of $\sigma$ from Eq.\ (\ref{counter}) into Eq.\ (\ref{Smarr3}) allows us to obtain
\vspace{-0.1cm}
\be X=1+\frac{4J_{H}^{2}}{\Big[M(R+2M)+|\mathcal{Q}|^{2}\Big]^{2}}, \ee

\vspace{-0.1cm}
\noi thus having the explicit form of $\sigma$
\vspace{-0.1cm}
\be \sigma=\sqrt{M^{2}-\left(|\mathcal{Q}|^{2}+\frac{J_{H}^{2}\big[(R+2M)^{2}+4|\mathcal{Q}|^{2}\big]}
{\big[M(R+2M)+|\mathcal{Q}|^{2}\big]^{2}}\right)\frac{R-2M}{R+2M}}. \label{thesigma}\ee

\vspace{-0.1cm}
Notice that there is a distinction between the horizon mass $M_{H}$ and the parameter $M$ as has been pointed out earlier by Cl\'ement and Gal'tsov \cite{Galtsov}; both are identical only whether the boundary term $M_{A}^{s}$ is killed due mainly to the fact that magnetic charges are not present. So, we have that in the limit $R \rightarrow \infty$, the horizon mass $M_{H}$ behaves as
\vspace{-0.1cm}
\begin{align} M_{H}&=\lim_{R  \rightarrow \infty}\Big[M-B_{H}^{2}\phi^{H}\Big]=M-B_{H}^{2} \frac{M+ \sigma}{(M+\sigma)^{2}+(J_{H}/M)^{2}},\nonu\\
\sigma&=\sqrt{M^{2}-|\mathcal{Q}|^{2}-J_{H}^{2}/M^{2}},\label{counterinfinite}
\end{align}

\vspace{-0.1cm}
\noi recovering the expression for one isolated dyonic Kerr-Newman BH joined to a DS \cite{Galtsov}. In this counterrotating sector, $J_{H}$ defines half of the total angular momentum of the system since it does not contain an extra contribution coming from the DS. On the other hand, the addition of the same constant $K_{0}=-B_{H}$ in the first corotating model with identical electric charges produces
\vspace{-0.1cm}
\begin{align}
M_{Anew}^{S}&= B_{H} \Bigg(\frac{B_{H} \epsilon \mathcal{L}}{\mathcal{L}^{2}+\mathcal{M}^{2}}+ \frac{2\mathfrak{q}Q_{H}}{P_{0}}\Bigg),\qquad
\Phi_{Enew}^{H}= \frac{Q_{H}(R+2\sigma)\mathcal{L}-(2b_{0}-B_{H}\epsilon)\mathcal{M}}{\mathcal{L}^{2}+\mathcal{M}^{2}}
+\frac{2\mathfrak{q}B_{H}}{P_{0}},\nonu\\
J_{H}&=2M\mathfrak{q}-\frac{\delta}{2}-Q_{H}B_{H},\label{ThermoDS1new}\end{align}

\vspace{-0.1cm}
\noi while its addition in the one carrying opposite electromagnetic charges generates the result
\vspace{-0.1cm}
\begin{align}
\qquad  M_{Anew}^{S}&= 2B_{H}\Bigg(  \frac{b_{o}\mathcal{N}}{\mathcal{N}^{2}+\mathcal{M}^{2}}-\frac{\mathfrak{q}Q_{H}}{P_{0}}\Bigg), \qquad 
\Phi_{Enew}^{H}= 2\Bigg(\frac{q_{o}\mathcal{N}}{\mathcal{N}^{2}+\mathcal{M}^{2}}+\frac{\mathfrak{q}B_{H}}{P_{0}}\Bigg), \nonu\\
J_{H}&=2M\mathfrak{q}-\frac{\delta}{2},
\label{ThermoDS2new}\end{align}

\vspace{-0.1cm}
\noi where it can be checked straightforwardly from these corotating models that Eq.\ (\ref{counterinfinite}) is recovered at an infinite separation distance, where $\mathfrak{q}=J_{H}/M$. The reader should further note that in the binary system with identical electric charges, $J_{H}$ equates the half of the total angular momentum only at large distances.

It should be mentioned here, that contrary to the statement provided by the authors in Ref.\ \cite{MankoCompean}, there is no need to add any constant to the magnetic potential $A_{3}$ when a DR procedure has been applied to include the magnetic charge inside the Kerr-Newman solution. These authors \cite{MankoCompean} have combined two approaches; the DR studied many years ago by Carter \cite{Carter} and the addition of a constant in the magnetic potential $A_{3}$,\footnote{The constant value agrees with $K_{0}=-B_{H}$ ($b_{0}=-\mathcal{B}$ in \cite{MankoCompean}). However, a constant must be added to $A_{3}$ only when the contribution of the DS into the horizon mass should be taken into account \cite{Galtsov}.} with the objective to prove the correctness of Tomimatsu's formulas in the presence of both electric and magnetic charges. However, their procedure cannot be correct since there was an error in such formulas that might lead to physical and mathematical inconsistencies. To prove the last statement on the no need of adding a constant in $A_{3}$, for simplicity and without loss of generality one might use once again the expressions of Eq.\ (\ref{counter})\footnote{These formulas were obtained in Ref.\ \cite{CLLM2} without adding a constant in the magnetic potential $A_{3}$.} after eliminating the magnetic charge, thus having
\vspace{-0.1cm}
\begin{align} M_{A}^{S}&=0, \quad \Phi_{E}^{H}=Q_{H}\phi^{H}, \nonu\\
\Omega&=\frac{\mu}{2}\frac{(R+2\sigma)\sqrt{X-1}}{M[R+2\sigma-(R-2M)X]-\mu Q_{H}^{2}},\qquad
\phi^{H}=\frac{\mu}{2}\frac{R+2\sigma-(R-2M)X}{M[R+2\sigma-(R-2M)X]-\mu Q_{H}^{2}},\nonu\\
\sigma&=\sqrt{X\big(M^{2}-Q_{H}^{2}\mu\big)+\frac{R^{2}}{4}(1-X)},\label{counter2}
\end{align}

\vspace{-0.1cm}
\noi where now the mass formula Eq.\ (\ref{Smarr}) reads
\vspace{-0.1cm}
\be M_{H} \equiv M=\sigma + 2\Omega J_{H} +Q_{H}^{2}\phi^{H}.\ee

\vspace{-0.1cm}
Then we have that a DR procedure ($Q_{H}\rightarrow Q_{H}+iB_{H}$ and $Q_{H}^{2}\rightarrow Q_{H}^{2}+B_{H}^{2}$) extends the conventional mass formula by adding the magnetic charge as a fourth conserved parameter as follows
\vspace{-0.1cm}
\begin{align} M_{H}&=\sigma + 2\Omega J_{H} +\Phi_{EL}^{H}Q_{H}+ \Phi_{MAG}^{H}B_{H},\nonu\\
\Phi_{EL}^{H}&=Q_{H}\phi^{H}, \qquad \Phi_{MAG}^{H}=B_{H}\phi^{H}, \label{Smarr4}\end{align}

\vspace{-0.1cm}
\noi with $\sigma$ having the same aspect as shown in Eq.\ (\ref{thesigma}), but the main distinction is that the parameter $M$ is now representing the horizon mass $M_{H}$. Hence, no DS exists in between BHs.

An heuristic point of view on the electromagnetic charge conservation is reached by killing first the magnetic charge with the condition \cite{Tomi}
\vspace{-0.1cm}
\be \frac{1}{4\pi }\int_{H}\omega A_{4,z}\, d\varphi dz=-\frac{1}{4\pi }\int_{H} A_{3,z}\, d\varphi dz=0, \label{condition}\ee

\vspace{-0.1cm}
\noi which means that both potentials $A_{4}$ and $A_{3}$ will take the same value in the limits defining the BH horizon. Therefore, the DS joined to the BH is eliminated, and later on, the DR $\Phi \rightarrow \Phi e^{i \alpha}$ is applied, where the real and imaginary components of the new potential $\Phi e^{i \alpha}$ take the aspect
\vspace{-0.1cm}
\be A_{4 new}=A_{4}+\frac{B_{H}}{Q_{H}} A_{3}^{'}, \qquad A_{3 new}^{'}=-\frac{B_{H}}{Q_{H}}A_{4}+A_{3}^{'}, \label{newones}\ee

\vspace{-0.1cm}
\noi whereby $\alpha=\arctan(B_{H}/Q_{H})$. Observe that the case $B_{H}=0$ recovers the original potential $\Phi$. Thus, we write the electric and magnetic charges as follows:
\vspace{-0.1cm}
\be Q_{Hnew}=\frac{1}{4\pi}\int_{H} \omega A_{3new,z}^{'}\, d\varphi dz, \qquad B_{Hnew}=\frac{1}{4\pi}\int_{H} \omega A_{4new,z}^{'}\, d\varphi dz, \label{newcharges}\ee

\vspace{-0.1cm}
\noi after the substitution of Eq.\ (\ref{newones}) into Eq.\ (\ref{newcharges}) it is possible to restore the contribution of the magnetic charge, namely
\vspace{-0.1cm}
\begin{align}  Q_{Hnew}&=\frac{1}{4\pi}\int_{H} \omega \left(-\frac{B_{H}}{Q_{H}}A_{4}+A_{3}^{'}\right)_{,z}\, d\varphi dz= Q_{H}, \nonu\\
B_{Hnew}&=\frac{1}{4\pi}\int_{H} \omega \left(A_{4}+\frac{B_{H}}{Q_{H}} A_{3}^{'}\right)_{,z}\, d\varphi dz = Q_{H} \left(\frac{B_{H}}{Q_{H}}\right)=B_{H}.
\label{thermo}\end{align}

\vspace{-0.1cm}
The same idea depicted above can be worked out in order to prove that $M_{H}$ and $J_{H}$ are also conserved parameters under a DR, where we have that the new horizon mass and angular momentum assume the form
\vspace{-0.1cm}
\begin{align}  M_{Hnew}&=M-M_{Anew}^{S}=M-M_{A}^{S}+\frac{B_{H}}{4\pi Q_{H}}\int_{H}(A_{4}A_{3})_{,z}d\varphi dz,\nonu\\
J_{Hnew}&=-\frac{1}{8\pi}\int_{H}\omega\left[1+ \frac{\omega \Psi_{,z}}{2}
-\tilde{A}_{3}A_{3new,z}^{'}\right]d\varphi dz -\frac{\omega^{H}M_{Anew}^{S}}{2},
\label{newmassandmomentum}
\end{align}

\vspace{-0.1cm}
\noi where $M_{Hnew}=M_{H}=M$ and $J_{Hnew}=J_{H}$ once Eq.\ (\ref{condition}) is fulfilled to eliminate the DS. We now turn our attention to consider a DR in the solution \cite{CCHV} in order to derive two corotating dyonic binary BH models with no DS in between.

\vspace{-0.4cm}
\subsection{Corotating dyonic BHs endowed with identical electromagnetic charges}
\vspace{-0.3cm}
An absence of magnetic charge ($B_{H}=0$) in Eq.\ (\ref{Sol1}) derives the following result
\vspace{-0.1cm}
\begin{align}  \delta&=\frac{2\mathfrak{q}(R^{2}-4\Delta)\big[MP_{0}+Q_{H}^{2}(R+2M)\big]}{(R^{2}+2MR+4\mathfrak{q}^{2})P_{0}
+8\mathfrak{q}^{2}Q_{H}^{2}},\qquad
b_{o}=-\frac{\mathfrak{q}Q_{H}(R^{2}-4\Delta)\big(P_{0}+2Q_{H}^{2}\big)}{(R^{2}+2MR+4\mathfrak{q}^{2})P_{0}
+8\mathfrak{q}^{2}Q_{H}^{2}},
\label{sol1}\end{align}

\vspace{-0.1cm}
\noi where the extra term $M_{A}^{S}$ has been eliminated and for such a reason $M_{H}=M$. In this respect, the DR can be performed by doing only the following changes $Q_{H}\rightarrow Q_{H}+iB_{H}$  and $Q_{H}^{2} \rightarrow Q_{H}^{2}+B_{H}^{2}$ in the Ernst potentials on the symmetry axis given above in Eq.\ (\ref{ernstaxiselectroI}). The result is
\vspace{-0.1cm}
\begin{align}
{\cal E}(0,z)&=\frac{z^{2}-2(M + i \mathfrak{q})z+2\Delta_{o}-R^{2}/4-\sigma^{2}+ i\delta}{z^{2}+2(M - i \mathfrak{q})z+2\Delta_{o}-R^{2}/4-\sigma^{2}- i\delta}, \qquad
\Phi(0,z)=\frac{2(\mathcal{Q} z+\mathfrak{q}_{o})}{z^{2}
+2(M - i \mathfrak{q})z+2\Delta_{o}-R^{2}/4-\sigma^{2}- i\delta},
\label{ernstaxiselectro}\end{align}

\vspace{-0.1cm}
\noi with
\vspace{-0.1cm}
\begin{align}
\delta&=\frac{2\mathfrak{q}(R^{2}-4\Delta_{o})\big[MP_{0}+|\mathcal{Q}|^{2}(R+2M)\big]}{(R^{2}+2MR+4\mathfrak{q}^{2})P_{0}
+8\mathfrak{q}^{2}|\mathcal{Q}|^{2}},\qquad
\mathfrak{q}_{o}= -\frac{i\mathfrak{q}\mathcal{Q}(R^{2}-4\Delta_{o})\big(P_{0}+2|\mathcal{Q}|^{2}\big)}{(R^{2}+2MR+4\mathfrak{q}^{2})P_{0}
+8\mathfrak{q}^{2}|\mathcal{Q}|^{2}},\nonu\\
\Delta_{o}&= M^{2}-|\mathcal{Q}|^{2}-\mathfrak{q}^{2}. \label{setvariables}\end{align}

\vspace{-0.1cm}
\noi It should be pointed out that this procedure has added the magnetic charge $B_{H}$ as a fourth conserved parameter into the solution, where now the electromagnetic charge obtainable from Eq.\ (\ref{totalcharge}) is given by
\vspace{-0.1cm}
\begin{align}
&Q_{H}+i B_{H}= Q_{H}+i B_{H}+2\frac{P_{0}(q_{o}+i b_{o})+i(Q_{H}+iB_{H})\Big(\mathfrak{q}(R^{2}-4\Delta_{o})+(R+2M)\delta\Big)}
{(R+2M)(R^{2}-4\Delta_{o})-4\mathfrak{q}\delta},
\label{totalchargeII}\end{align}

\vspace{-0.1cm}
\noi which is identically satisfied by the set of variables expressed lines above in Eq.\ (\ref{setvariables}). Similar to the upper dyonic BH, the lower constituent contains the same electromagnetic charge $Q_{H}+iB_{H}$. Once we have incorporated the magnetic charge $B_{H}$, it follows that the Ernst potentials, Kinnersley potential $\Phi_{2}$, and metric functions read
\vspace{-0.1cm}
\begin{align}
{\cal{E}}&=\frac{\Lambda+\Gamma}{\Lambda-\Gamma},\quad \Phi=\frac{\chi}{\Lambda-\Gamma},\qquad \Phi_{2}=\frac{F}{\Lambda-\Gamma},\qquad f=\frac{|\Lambda|^{2}-|\Gamma|^{2}+ |\chi|^{2}}{|\Lambda-\Gamma|^{2}},\qquad \omega=4\mathfrak{q}+\frac{{\rm{Im}}\left[(\Lambda-\Gamma)\overline{\mathcal{G}}-\chi \overline{\mathcal{I}} \right]}{|\Lambda|^{2}-|\Gamma|^{2}+ |\chi|^{2}},\nonu\\
e^{2\gamma}&=\frac{|\Lambda|^{2}-|\Gamma|^{2}+ |\chi|^{2}}{64\sigma^{4}R^{4}\kappa_{o}^{2} r_{1}r_{2}r_{3}r_{4}}, \qquad \Lambda=2\sigma^{2} \left[R^{2}\kappa_{o}(r_{1}+r_{2})(r_{3}+r_{4})+4a(r_{1}-r_{3})(r_{2}-r_{4})\right]\nonu\\
&+2R^{2}\left[\kappa_{o}(2\Delta_{o}-\sigma^{2})-a\right](r_{1}-r_{2})(r_{3}-r_{4}) +2iR\bigg\{\Big(2\mathfrak{q}{\rm Re}(s_{+})+{\rm Im}(p_{+})\Big)\Big[R(\mathfrak{r}_{1}-\mathfrak{r}_{2})(r_{3}-r_{4})\nonu\\
&-2\sigma\big(\mathfrak{r}_{1}r_{4}-\mathfrak{r}_{2}r_{3}+4\sigma r_{3}r_{4}\big)\Big]
+\mathfrak{q}\kappa_{o}\Big[ r_{1}\big( R^{2}r_{3}-\kappa_{o}r_{4}\big)-r_{2}\big(\kappa_{o} r_{3}-R^{2}r_{4}\big)-8\sigma^{2}r_{3}r_{4} \Big]\bigg\},\nonu\\
\Gamma&=4\sigma R\left(M\Gamma_{o}- b\chi_{+}\right),\qquad F=(4\mathfrak{q}+iz)\chi-i\mathcal{I},\qquad \chi=-4\sigma R\left(\mathcal{Q}\Gamma_{o}+2\mathbb{Q}\chi_{+}\right),\qquad \Gamma_{o}=R \chi_{-}-2\sigma \chi_{s}+2\chi_{1+}, \nonu\\
\mathcal{G}&= 2z\Gamma+  8\sigma^{2}\bigg\{ R\Big[2\big({\rm Re}(a)-2|\mathfrak{q}_{o}|^{2}\big)
+|\mathcal{Q}|^{2}\kappa_{o}\Big](r_{1}r_{2}-r_{3}r_{4})+2i\mathfrak{q}R^{2}\kappa_{o}(r_{2}r_{3}+r_{1}r_{4})
+2i\Big[R{\rm Im}(a)+\overline{\mathcal{Q}}\xi_{0}-4\mathfrak{q}|\mathfrak{q}_{o}|^{2}\Big]\nonu\\
&\times(r_{1}-r_{3})(r_{2}-r_{4})\bigg\}
-4R^{2}\bigg\{\sigma\left[2a-(R-2\sigma)\Big(2(R+2i\mathfrak{q})s_{+}+p_{+} \Big)\right]+i\left(\overline{\mathcal{Q}}\xi_{o}+2i \mathfrak{q}_{o}\overline{\mathcal{Q}}\kappa_{o}-4\mathfrak{q}|\mathfrak{q}_{o}|^{2}\right)\bigg\}
(r_{1}-r_{2})(r_{3}-r_{4}) \nonu\\
&+2\sigma R^{2}\bigg\{4\Big(2\kappa_{o} \Delta_{o} -{\rm Re}(a)\Big)r_{4}  +\big[|\mathcal{Q}|^{2}\kappa_{o}+4|\mathfrak{q}_{o}|^{2}\big](r_{3}+r_{4})\bigg\}(r_{1}-r_{2}) \nonu\\
&+2\sigma R^{2} \bigg\{4\Big(2\kappa_{o} \Delta_{o} -{\rm Re}(a)\Big)r_{2} +\big[|\mathcal{Q}|^{2}\kappa_{o}+4|\mathfrak{q}_{o}|^{2}\big](r_{1}+r_{2})\bigg\}(r_{3}-r_{4})
+4M\sigma R\big(\kappa_{o}\chi_{+} +2R\chi_{1-}+4\sigma^{2}\chi_{p}\big) \nonu\\
&-4b\sigma R(R\chi_{-}+2\sigma\chi_{s})-8\sigma R(\mathcal{Q}b+2M\mathbb{Q})\Big[ 2\overline{\mathfrak{q}}_{o}\big(\mathfrak{r}_{1}-\mathfrak{r}_{2}+\mathfrak{r}_{3}-\mathfrak{r}_{4}\big)
+\overline{\mathcal{Q}}\kappa_{o}(r_{1}-r_{2}-r_{3}+r_{4})\Big],\nonu\\
\mathcal{I}&=A\Big[4\sigma^{2}(r_{1}-r_{3})(r_{2}-r_{4})-R^{2}(r_{1}-r_{2})(r_{3}-r_{4})\Big]
+R\kappa_{-}\Big[B_{+}\kappa_{o}r_{1}-B_{-}R\mathfrak{r}_{2} \Big]r_{4}-R\kappa_{+}\Big[B_{+}R\mathfrak{r}_{1}-B_{-}\kappa_{o}r_{2} \Big]r_{3}\nonu\\
&-16\sigma^{2}R \bigg\{\Big[M(R+2\sigma)(\kappa_{+}+2\mathcal{Q}R)-B_{+}\mathfrak{q}_{o}\Big]r_{3}r_{4}
-R\kappa_{o}(2M\mathbb{Q}+\mathcal{Q}b)\bigg\}+8\mathbb{Q}\sigma R(\chi_{1+}+\sigma\chi_{s})\nonu\\
&+ 2\sigma R\Big[\mathcal{Q}\big(2R^{2}-8\Delta_{o}+\kappa_{o}\big)+8i\mathfrak{q} \mathbb{Q}\Big]\chi_{+}+ 12\sigma R^{2}\mathbb{Q}\chi_{-} + 8\mathcal{Q}\sigma R(R\chi_{1-}+2\sigma\chi_{p}), \nonu\\
\chi_{\pm}&=s_{+}\mathfrak{r}_{1}-s_{-}\mathfrak{r}_{2} \pm
(\bar{s}_{-}\mathfrak{r}_{3}-\bar{s}_{+}\mathfrak{r}_{4}), \quad \chi_{1\pm}=p_{+}\mathfrak{r}_{1}+p_{-}\mathfrak{r}_{2}\pm
(\bar{p}_{-}\mathfrak{r}_{3}+\bar{p}_{+}\mathfrak{r}_{4}),\quad
\chi_{s}=s_{+}\mathfrak{r}_{1}+s_{-}\mathfrak{r}_{2}+\bar{s}_{-}\mathfrak{r}_{3}+\bar{s}_{+}\mathfrak{r}_{4}, \nonu\\
\chi_{p}&=p_{+}\mathfrak{r}_{1}-p_{-}\mathfrak{r}_{2} + \bar{p}_{-}\mathfrak{r}_{3}-\bar{p}_{+}\mathfrak{r}_{4},\quad a=(R+2i\mathfrak{q})p_{+}-s_{+}\big[s_{+}-(R+2i\mathfrak{q})^{2}\big], \quad
b=i(\delta-4M\mathfrak{q}), \quad \kappa_{o}=R^{2}-4\sigma^{2},\nonu\\
A&=4M\Big[ \big( 2 \mathbb{Q}+ \mathcal{Q}(R-2\sigma)\big)s_{+}+2\mathcal{Q}p_{+}\Big]
+B_{+}\Big[\mathcal{Q}\big(R^{2}-4\Delta_{o}\big)-2(R+2i\mathfrak{q})\mathfrak{q}_{o} \Big], \quad
\kappa_{\pm}=2\mathfrak{q}_{o}-\mathcal{Q}(R\pm 2\sigma),\nonu\\
B_{\pm}&=\Big[ R s_{\pm} \pm p_{\pm} +2\mathcal{Q}\big(2\overline{\mathfrak{q}}_{o}+\overline{\mathcal{Q}}(R\pm 2\sigma)\big)\Big]/M, \quad
p_{\pm}=-\sigma(R^{2}-4\Delta_{o})\pm i\big[2M\delta+4i \mathcal{Q} \overline{\mathfrak{q}}_{o}-(R+2i\mathfrak{q}){\rm Im}(s_{\pm})\big],\nonu\\ 
s_{\pm}&=2\Delta_{o}\pm \sigma R+ i \mathfrak{q}(R\pm2\sigma), \quad
\xi_{o}=4\mathcal{Q}\Big[M \delta-2i \mathfrak{q}_{o} \overline{\mathcal{Q}}+\mathfrak{q}(\Delta_{o}-\sigma^{2})\Big]+(2i \mathfrak{q}_{o}-\mathfrak{q} \mathcal{Q})(R^{2}-4\Delta_{o}), \quad \mathbb{Q}=\mathfrak{q}_{o}+2i \mathfrak{q} \mathcal{Q}, \nonu\\
\mathfrak{r}_{1,4}&=(R-2\sigma)r_{1,4}, \quad \mathfrak{r}_{2,3}=(R+2\sigma)r_{2,3},
\label{fiveparameters}
\end{align}

\vspace{-0.1cm}
\noi where the half-length parameter defining the BH horizon can be written as
\vspace{-0.1cm}
\be \sigma= \sqrt{\Delta_{o}- \frac{4|\mathfrak{q}_{o}|^{2}-\delta^{2}}{R^{2}-4\Delta_{o}}},\label{equalsigma}\ee

\vspace{-0.1cm}
\noi which explicitly is
\vspace{-0.1cm}
\begin{align}  \sigma&=\sqrt{\Delta_{o} +\frac{4\mathfrak{q}^{2}(R^{2}-4\Delta_{o})\Big[\big[MP_{0}+|\mathcal{Q}|^{2}(R+2M)\big]^{2}
-|\mathcal{Q}|^{2}(P_{0}+2|\mathcal{Q}|^{2})^{2}\Big]}
{\big[(R^{2}+2MR+4\mathfrak{q}^{2})P_{0}+8\mathfrak{q}^{2}|\mathcal{Q}|^{2}\big]^{2}}}.
\label{sigmamod}\end{align}

\vspace{-0.1cm}
In this case the components of the extended mass formula Eq.\ (\ref{Smarr4}), $\Omega$ and $\phi^{H}$ are given by
\vspace{-0.1cm}
\begin{align}  \Omega&=\frac{2\mathfrak{q}}{P_{0}}+ \frac{P_{0}R(R^{2}-4\Delta_{o})}{(R^{2}+2MR+4\mathfrak{q}^{2})P_{0}
+8\mathfrak{q}^{2}|\mathcal{Q}|^{2}}\Bigg(\frac{\mathcal{M}}{\mathcal{L}_{o}^{2}+\mathcal{M}^{2}}\Bigg), \qquad
\phi^{H}=\frac{(R+2\sigma)\mathcal{L}_{o}-2(b_{o}/Q_{H})\mathcal{M}}
{\mathcal{L}^{2}_{o}+\mathcal{M}^{2}},\nonu\\
\mathcal{L}_{o}&=MR+2\Delta_{o}+(R+2M)\sigma,
\label{thermo}\end{align}

\vspace{-0.1cm}
\noi and their combination with Eq.\. (\ref{sigmamod}) defines the angular momentum from Eq.\ (\ref{Smarr4}), to obtain
\vspace{-0.1cm}
\be J_{H}= 2M\mathfrak{q}
-\frac{\mathfrak{q}(R^{2}-4\Delta_{o})\big[MP_{0}+|\mathcal{Q}|^{2}(R+2M)\big]}{(R^{2}+2MR+4\mathfrak{q}^{2})P_{0}
+8\mathfrak{q}^{2}|\mathcal{Q}|^{2}}, \label{momentum}\ee

\vspace{-0.1cm}
\noi which is nothing less than half of the total angular momentum;i.e., $2M\mathfrak{q}-\delta/2$. In addition, we have that the area of the horizon $S$ and surface gravity $\kappa$ are expressed as
\vspace{-0.1cm}
\be \frac{S}{4\pi}= \frac{\sigma}{\kappa}=\frac{\mathcal{L}_{o}^{2}+\mathcal{M}^{2}}{R(R+2\sigma)}.\ee

\vspace{-0.1cm}
The conical singularity (or strut) along the line keeping apart the BHs from overlapping presents an angular deficit $\Delta \varphi= -8\pi \mathcal{\mathcal{F}}$, where the interaction force $\mathcal{F}$ is computed with the aid of the formula $\mathcal{F}=(e^{-\gamma_{s}}-1)/4$ \cite{Israel,Weinstein}, being $\gamma_{s}$ the constant value for the metric function $\gamma$ in the axis region in between sources. The result is
\vspace{-0.1cm}
\begin{align}  \mathcal{F}&= \frac{\big[(M^{2}-|\mathcal{Q}|^{2})P_{0}^{2}-4\mathfrak{q}^{2}|\mathcal{Q}|^{4}\big](P_{0}-8\mathfrak{q}^{2})
-16\mathfrak{q}^{2}|\mathcal{Q}|^{2}
\Big[\big(M(R+2M)-|\mathcal{Q}|^{2}\big)P_{0}-|\mathcal{Q}|^{4}\Big]}{(R^{2}-4\Delta_{o})P_{0}^{3}}.\label{theforce1}\end{align}

\vspace{-0.1cm}
The strut is massless because does not contribute to the total gravitational energy of the system. The force $\mathcal{F} \rightarrow 0$ at $R \rightarrow \infty$, where Eqs.\ (\ref{sigmamod}) and (\ref{momentum}) permit us to recover the expression $\sigma=\sqrt{M_{H}^{2}-|\mathcal{Q}|^{2}-J_{H}^{2}/M_{H}^{2}}$ defining an isolated dyonic BH free of monopolar sources. Moreover, not taking into account large separation distances, the lack of a conical singularity is achieved when $\mathcal{F}=0$, and due to the fact that the numerator of $\mathcal{F}$ is a bicubic equation in the variable $\mathfrak{q}$, in principle this task can be done analytically. This matter clearly deserves further research in order to understand better the presence of ring singularities off the axis or other pathologies like closed time-like curves.

We end this subsection, by mentioning that the aforementioned expression of the force Eq.\ (\ref{theforce1}) will contain a different aspect if the DS is taken into account in the solution.

\vspace{-0.4cm}
\subsection{Corotating dyonic BHs endowed with opposite electromagnetic charges}
\vspace{-0.3cm}
The second charged model arises after setting $B_{H}=0$ in Eq.\ (\ref{Sol2}), thus having
\vspace{-0.1cm}
\begin{align}  \delta&=\frac{2\mathfrak{q}(R^{2}-4\Delta_{1})\big[MP_{0}-Q_{H}^{2}(R+2M)\big]}
{(R^{2}+2MR+4\mathfrak{q}^{2})P_{0}
-8\mathfrak{q}^{2}Q_{H}^{2}},\qquad
q_{o}=\frac{Q_{H}R(R^{2}-4\Delta_{1})P_{0}}{2\Big[(R^{2}+2MR+4\mathfrak{q}^{2})P_{0}
-8\mathfrak{q}^{2}Q_{H}^{2}\Big]},
\label{sol2}\end{align}

\vspace{-0.1cm}
\noi where now each BH contains an opposite electric charge. So, in this case the Ernst potentials on the symmetry axis after performing the DR are written as
\vspace{-0.1cm}
\begin{align}
{\cal E}(0,z)&=\frac{z^{2}-2(M + i \mathfrak{q})z+2\Delta_{1}-R^{2}/4-\sigma^{2}+ i\delta}{z^{2}+2(M - i \mathfrak{q})z+2\Delta_{1}-R^{2}/4-\sigma^{2}- i\delta}, \qquad
\Phi(0,z)=\frac{2\mathfrak{q}_{o}}{z^{2}
+2(M - i \mathfrak{q})z+2\Delta_{1}-R^{2}/4-\sigma^{2}- i\delta},
\label{ernstaxiselectro2}\end{align}

\vspace{-0.1cm}
\noi with $\delta$ and $q_{o}$ having the same aspect as in Eq.\ (\ref{Sol2}), but now $M=M_{H}$ since $M_{A}^{S}=0$.
The electromagnetic charge derived from Eq.\ (\ref{totalcharge}) and satisfied by Eq.\ (\ref{Sol2}) reduces to
\vspace{-0.1cm}
\be Q_{H}+i B_{H}=\frac{2(q_{o}+ib_{o})P_{0}}
{(R+2M)(R^{2}-4\Delta_{1})-4\mathfrak{q} \delta}. \label{totalchargeIII}\ee

\vspace{-0.1cm}
It is worthwhile to mention, that the lower dyonic BH is endowed with opposite electromagnetic charge; i.e.,$-Q_{H}-iB_{H}$. Then, the Ernst and Kinnersley potentials, as well as the metric functions, have now the form

\vspace{-0.1cm}
\begin{align}
{\cal{E}}&=\frac{\Lambda+\Gamma}{\Lambda-\Gamma},\qquad \Phi=\frac{\chi}{\Lambda-\Gamma},\qquad \Phi_{2}=\frac{F}{\Lambda-\Gamma},\qquad
f=\frac{|\Lambda|^{2}-|\Gamma|^{2}+ |\chi|^{2}}{|\Lambda-\Gamma|^{2}},\qquad \omega=4\mathfrak{q}+\frac{{\rm{Im}}\left[(\Lambda-\Gamma)\overline{\mathcal{G}}-\chi \overline{\mathcal{I}} \right]}{|\Lambda|^{2}-|\Gamma|^{2}+ |\chi|^{2}}, \nonu\\
e^{2\gamma}&=\frac{|\Lambda|^{2}-|\Gamma|^{2}+ |\chi|^{2}}{64\sigma^{4}R^{4}\kappa_{o}^{2} r_{1}r_{2}r_{3}r_{4}}, \qquad
\Lambda=2\sigma^{2} \left[R^{2}\kappa_{o}(r_{1}+r_{2})(r_{3}+r_{4})+4a(r_{1}-r_{3})(r_{2}-r_{4})\right]\nonu\\ &+2R^{2}\left[\kappa_{o}(2\Delta_{1}-\sigma^{2})-a\right](r_{1}-r_{2})(r_{3}-r_{4})
+2iR\bigg\{\Big(2\mathfrak{q}{\rm Re}(s_{+})+{\rm Im}(p_{+})\Big)\Big[R(\mathfrak{r}_{1}-\mathfrak{r}_{2})(r_{3}-r_{4})\nonu\\
&-2\sigma\big(\mathfrak{r}_{1}r_{4}-\mathfrak{r}_{2}r_{3}+4\sigma r_{3}r_{4}\big)\Big]
+\mathfrak{q}\kappa_{o}\Big[ r_{1}\big( R^{2}r_{3}-\kappa_{o}r_{4}\big)-r_{2}\big(\kappa_{o} r_{3}-R^{2}r_{4}\big)-8\sigma^{2}r_{3}r_{4} \Big]\bigg\},\nonu\\
\Gamma&=4\sigma R\left(M\Gamma_{o}- b\chi_{+}\right),\quad F=(4\mathfrak{q}+iz)\chi-i\mathcal{I},
\quad \chi=-8\sigma R \mathfrak{q}_{o}\chi_{+},\quad \Gamma_{o}=R \chi_{-}-2\sigma \chi_{s}+2\chi_{1+}, \nonu\\
\mathcal{G}&= 2z\Gamma+  8\sigma^{2}\Bigg\{ 2R\big({\rm Re}(a)-2|\mathfrak{q}_{o}|^{2}\big)
(r_{1}r_{2}-r_{3}r_{4})+2i\mathfrak{q}R^{2}\kappa_{o}(r_{2}r_{3}+r_{1}r_{4})+2i\Big[R{\rm Im}(a)-4\mathfrak{q}|\mathfrak{q}_{o}|^{2}\Big](r_{1}-r_{3})(r_{2}-r_{4}) \nonu\\
&-4R^{2}\bigg(\sigma\left[2a-(R-2\sigma)
\Big(2(R+2i\mathfrak{q})s_{+}+p_{+} \Big)\right]-4i\mathfrak{q}|\mathfrak{q}_{o}|^{2}\bigg)(r_{1}-r_{2})(r_{3}-r_{4})\Bigg\}\nonu\\
&+8\sigma R^{2}\Big[\big(2\kappa_{o} \Delta_{1} -{\rm Re}(a)\big)r_{4}+|\mathfrak{q}_{o}|^{2}(r_{3}+r_{4})\Big](r_{1}-r_{2}) +8\sigma R^{2} \Big[\big(2\kappa_{o} \Delta_{1} -{\rm Re}(a)\big)r_{2}+|\mathfrak{q}_{o}|^{2}(r_{1}+r_{2})\Big](r_{3}-r_{4})\nonu\\
&+4\sigma R \bigg\{M\Big[\kappa_{o}\chi_{+}
+2R\chi_{1-}+4\sigma\chi_{p}-8|\mathfrak{q}_{o}|^{2}(\mathfrak{r}_{1}-
\mathfrak{r}_{2}+\mathfrak{r}_{3}-\mathfrak{r}_{4})\Big]-b\big(R\chi_{-}+2\sigma\chi_{s}\big)\bigg\},\nonu\\
\mathcal{I}&=4\mathfrak{q}_{o}\bigg\{A\Big[4\sigma^{2}(r_{1}-r_{3})(r_{2}-r_{4})-R^{2}(r_{1}-r_{2})(r_{3}-r_{4})\Big]+
R\Big[B_{+}\Big((R+2\sigma)r_{4}-R r_{3}\Big)\mathfrak{r}_{1}+B_{-}\Big((R-2\sigma)r_{3}-R r_{4}\Big)\mathfrak{r}_{2}\Big]\nonu\\
&-8\sigma^{2}R\Big[\Big(M(R+2\sigma)-B_{+}\Big)r_{3}r_{4}-MR k_{o}\Big] +2\sigma R(\chi_{1+}+\sigma \chi_{s})+4i\mathfrak{q}\sigma R \chi_{+}+3\sigma R^{2} \chi_{-}\bigg\}, \nonu\\ 
\chi_{\pm}&=s_{+}\mathfrak{r}_{1}-s_{-}\mathfrak{r}_{2} \pm
(\bar{s}_{-}\mathfrak{r}_{3}-\bar{s}_{+}\mathfrak{r}_{4}), \quad
\chi_{1\pm}=p_{+}\mathfrak{r}_{1}+p_{-}\mathfrak{r}_{2}\pm
(\bar{p}_{-}\mathfrak{r}_{3}+\bar{p}_{+}\mathfrak{r}_{4}),\quad
\chi_{s}=s_{+}\mathfrak{r}_{1}+s_{-}\mathfrak{r}_{2}+\bar{s}_{-}\mathfrak{r}_{3}+\bar{s}_{+}\mathfrak{r}_{4}, \nonu\\
\chi_{p}&=p_{+}\mathfrak{r}_{1}-p_{-}\mathfrak{r}_{2} + \bar{p}_{-}\mathfrak{r}_{3}-\bar{p}_{+}\mathfrak{r}_{4},
\quad a=(R+2i\mathfrak{q})p_{+}-s_{+}\big[s_{+}-(R+2i\mathfrak{q})^{2}\big], \quad 
 b=i(\delta-4M\mathfrak{q}), \quad \kappa_{o}=R^{2}-4\sigma^{2}, \nonu\\
A&=2Ms_{+}-(R+2i\mathfrak{q})B_{+}, \quad  B_{\pm}=M(R\pm 2\sigma)+i\delta, \quad  p_{\pm}=-\sigma(R^{2}-4\Delta_{1})\pm i\big[2M\delta-(R+2i\mathfrak{q}){\rm Im}(s_{\pm})\big], \nonu\\
s_{\pm}&=2\Delta_{1}\pm \sigma R+i \mathfrak{q}(R\pm2\sigma),\quad
\mathfrak{r}_{1,4}=(R-2\sigma)r_{1,4}, \quad \mathfrak{r}_{2,3}=(R+2\sigma)r_{2,3},
 \quad 
 \label{opposite}\end{align}

\vspace{-0.1cm}
\noi with $\sigma$ having now the aspect
\vspace{-0.1cm}
\begin{align}  \sigma&=\sqrt{\Delta_{1} +\frac{(R^{2}-4\Delta_{1})\Big[\big[2\mathfrak{q}\big(MP_{0}-|\mathcal{Q}|^{2}(R+2M)\big)\big]^{2}
-(|\mathcal{Q}|RP_{0})^{2}\Big]}
{\big[(R^{2}+2MR+4\mathfrak{q}^{2})P_{0}-8\mathfrak{q}^{2}|\mathcal{Q}|^{2}\big]^{2}}}.
\label{sigma2}\end{align}

\vspace{-0.1cm}
On the other hand, the thermodynamical properties of the extended Smarr formula Eq.\ (\ref{Smarr4}) become
\vspace{-0.1cm}
\begin{align}  \Omega&=\frac{2\mathfrak{q}}{P_{0}}+ \phi^{H}, \qquad \phi^{H}=\frac{2(q_{o}/Q_{H})\mathcal{N}}
{\mathcal{N}^{2}+\mathcal{M}^{2}}, \qquad \frac{S}{4\pi}= \frac{\sigma}{\kappa}=\frac{\mathcal{N}^{2}+\mathcal{M}^{2}}{R(R+2\sigma)},
\label{thermoII}\end{align}

\vspace{-0.1cm}
\noi where they define the angular momentum of the horizon in the way
\vspace{-0.1cm}
\be J_{H}=2M\mathfrak{q}-\frac{\mathfrak{q}(R^{2}-4\Delta_{1})\big[MP_{0}-|\mathcal{Q}|^{2}(R+2M)\big]}
{(R^{2}+2MR+4\mathfrak{q}^{2})P_{0}
-8\mathfrak{q}^{2}|\mathcal{Q}|^{2}}. \label{angularmomentumopp}\ee

\vspace{-0.1cm}
Moreover, it is not difficult to show that the interaction force is given by
\vspace{-0.1cm}
\be\mathcal{F}= \frac{\big(M^{2}P_{0}^{2}-4\mathfrak{q}^{2}|\mathcal{Q}|^{4}\big)(P_{0}-8\mathfrak{q}^{2})+|\mathcal{Q}|^{2}
\big[R^{2}P_{0}-4\mathfrak{q}^{2}(R^{2}-4\Delta_{1})\big] P_{0}}{(R^{2}-4\Delta_{1})P_{0}^{3}},
\label{force2}\ee

\vspace{-0.1cm}
It is noteworthy that if the DS is not absent in the solution, the interaction force looks the same as in Eq.\ (\ref{force2}), nonetheless $M \neq M_{H}$. In the same manner as in the first case, the strut may be removed by solving another bicubic equation in terms of $\mathfrak{q}$. To finalize the section, we mention that if $B_{H}=0$, all the physical and thermodynamical features in both models are reduced to those ones defining corotating binary systems of identical Kerr-Newman BHs \cite{CCHV}.

\vspace{-0.4cm}
\section{Concluding remarks}
\vspace{-0.3cm}
Following Carter's approach \cite{Carter}, we have been able to apply a DR in two identical corotating Kerr-Newman binary BH models recently studied in \cite{CCHV} with the purpose to add individual magnetic charges to each BH. Therefore, each corotating BH is endowed with identical/opposite electromagnetic charge in the first/second configuration and satisfying a generalized Smarr formula for dyonic BHs. These models containing a conical singularity in between sources, are well represented by five physical arbitrary parameters $\{M_{H},J_{H},Q_{H},B_{H},R\}$ and they will be useful to provide further analytical studies of some astrophysical phenomena like geodesics, quasinormal modes or lensing and shadow effects in the context of binary BHs, among others. It is worth remarking that the DR approach describes configurations free of DS joined to the BHs. On the other hand, we have also shown in both corotating dyonic models that the mass $M_{H}$, angular momentum $J_{H}$, and electromagnetic charge $Q_{H}+ i B_{H}$ are conserved parameters under DR and there is no necessity to add a gauge in the magnetic potential $A_{3}$ as has been claimed in \cite{MankoCompean}. On the contrary, the addition of a gauge is only needed when the contribution of the DS into the horizon mass $M_{H}$ is introduced in order to balance each dyonic BH. In all the scenarios examined in this work, this gauge is a constant of value $K_{0}=-B_{H}$ that leads us to the correct description of an isolated BH joined to a DS previously studied by Cl\'ement and Gal'tsov \cite{Galtsov}.

We would like to point out that the metric for the extreme limit case $(\sigma=0)$ of corotating dyonic BHs has not been considered here since it can be easily derived from the formulas given in \cite{CCHV}, where both electrically charged configurations were well defined. In fact, it should be mentioned that both dyonic configurations satisfy the Gabach-Clement identity \cite{Maria} for extreme BHs with struts, namely
\vspace{-0.1cm}
\be \sqrt{1+4\mathcal{F}_{ext}}=\frac{\sqrt{(8\pi J_{H})^{2}+(4\pi Q_{H}^{2}+4\pi B_{H}^{2})^{2}}}{S_{ext}}, \label{GabachClement}\ee

\vspace{-0.1cm}
\noi where $S_{ext}$ and $\mathcal{F}_{ext}$ define the horizon area and the force in the extreme case, respectively. Due to the fact that the expression of the force contains the same aspect in both extreme and non-extreme configurations, because it does not depend on $\sigma$. For instance, in the identically electromagnetic charged model, we have that
\vspace{-0.1cm}
\be S_{ext}=4\pi\frac{(MR+2\Delta_{o})^{2}+(\delta+\mathfrak{q} R)^{2}}{R^{2}}, \label{areaextreme1}\ee

\vspace{-0.1cm}
\noi and the substitution of Eqs.\ (\ref{momentum}), (\ref{theforce1}), and (\ref{areaextreme1}) into Eq.\ (\ref{GabachClement}) yields
\vspace{-0.1cm}
\be \Delta_{o} +\frac{4\mathfrak{q}^{2}(R^{2}-4\Delta_{o})\Big[\big[MP_{0}+|\mathcal{Q}|^{2}(R+2M)\big]^{2}
-|\mathcal{Q}|^{2}(P_{0}+2|\mathcal{Q}|^{2})^{2}\Big]}
{\big[(R^{2}+2MR+4\mathfrak{q}^{2})P_{0}+8\mathfrak{q}^{2}|\mathcal{Q}|^{2}\big]^{2}}=0,
\label{conditionClement}\ee

\noi which is exactly the condition $\sigma=0$ on Eq.\ (\ref{sigmamod}) for extreme dyonic BHs. Furthermore, one may proceed in the same manner for the oppositely dyonic charged configuration in order to derive once again the expression that arises from the condition $\sigma=0$ imposed on Eq.\ (\ref{sigma2}).

\vspace{-0.4cm}
\section*{Acknowledgements}
\vspace{-0.3cm}
This paper is dedicated to the memory of Prof. Oscar Figueroa Cruz. The author thanks the referee for his valuable remarks and suggestions. The financial support of SNI-CONACyT, M\'exico, grant with CVU No. 173252 is also acknowledged.

\end{document}